\documentclass[prb,twocolumn,superscriptaddress,showpacs,
 amsmath,amssymb]{revtex4}

\usepackage{graphicx} % Include figure files
\usepackage{dcolumn} % Align table columns on decimal point
\usepackage{hyperref}

\def\bk{{\bf k}}

\def\bQ{{\bf Q}}

\def\b0{{\bf 0}}

\def\bra{\langle}
\def\ket{\rangle}
\def\up{\uparrow}
\def\down{\downarrow}

\def\eps{\epsilon}

\def\Gam{\Gamma}

\def\Lam{\Lambda}

\def\sg{\sigma}
\def\Sg{\Sigma}

\begin{document}

\title{Competing order in correlated electron systems made simple: Consistent 
fusion of functional renormalization and mean-field theory}

\author{Jing Wang}
\affiliation{Max Planck Institute for Solid State Research,
 D-70569 Stuttgart, Germany}
\affiliation{Department of Modern Physics, University of Science
 and Technology of China, Hefei, Anhui 230026, P.~R.~China}
\author{Andreas Eberlein}
\affiliation{Max Planck Institute for Solid State Research,
 D-70569 Stuttgart, Germany}
\author{Walter Metzner}
\affiliation{Max Planck Institute for Solid State Research,
 D-70569 Stuttgart, Germany}

\date{\today}

\begin{abstract}
We derive an efficient and unbiased method for computing order 
parameters in correlated electron systems with competing 
instabilities. 
Charge, magnetic and pairing fluctuations above the energy scale of 
spontaneous symmetry breaking are taken into account by a functional
renormalization group flow, while the formation of order below 
that scale is treated in mean-field theory.
The method captures fluctuation driven instabilities such as
$d$-wave superconductivity.
As a first application we study the competition between 
antiferromagnetism and superconductivity in the ground state of
the two-dimensional Hubbard model.
\end{abstract}
\pacs{71.10.Fd, 74.20.-z, 75.10.-b}

\maketitle

%%% Intro %%%%%%%%%%%%%%%%%%%%%%%%%%%%%%%%%%%%%%%%%%%%%%%%%%%%%%%%

Competing order is a ubiquitous phenomenon in two-dimensional
interacting electron systems. A most prominent example is the 
competition between antiferromagnetism and high temperature 
superconductivity in cuprate and iron pnictide compounds.
Some of the ordering tendencies are fluctuation driven, and 
can therefore not be captured by mean-field (MF) theory.
Numerical simulations of correlated electrons are still
restricted to relatively small systems.

For weak and moderate interaction strengths, the functional
renormalization group (fRG) has been developed as an unbiased
and sensitive tool to detect instabilities toward any kind of
order in interacting electron models.\cite{metzner12}
In that method, effective interactions, self-energies, and
susceptibilities are computed from a differential flow equation,
where the flow parameter $\Lam$ controls a scale-by-scale 
integration of fields in the underlying functional integral.
Instabilities are signalled by divergences of effective 
interactions and susceptibilities at a critical energy scale
$\Lam_c$.
To complete the calculation and compute, for example, the size
of the order parameters, one has to continue the flow below 
the scale $\Lam_c$, which requires the implementation of 
spontaneous symmetry breaking.
This can be done either in a purely fermionic framework
\cite{salmhofer04} or by introducing bosonic order parameter
fields.\cite{baier04}
Both approaches have been applied already to interacting
electron models, such as the two-dimensional Hubbard model 
with repulsive \cite{baier04,friederich10,eberlein13a} and
attractive \cite{strack08,gersch08,eberlein13} interactions.

The flow in the symmetry-broken regime ($\Lam < \Lam_c$) 
is complicated considerably by the presence of anomalous 
interaction vertices.
In complex problems, such as systems with several competing
and possibly coexisting order parameters, or in multi-band
systems, it can therefore be mandatory or at least desirable
to simplify the integration of the scales below $\Lam_c$.
A natural possibility is to treat the low-energy degrees
of freedom (below $\Lam_c$) in mean-field theory.
The generation of instabilities and also the possible 
reduction of the critical scale by fluctuations is not 
affected by such a simplification.
In the {\em ground state}, fluctuations below the critical
scale are expected to influence the size of order parameters
only mildly. This has been confirmed for the attractive and
repulsive Hubbard model by several previous fRG studies.
\cite{baier04,strack08,eberlein13,eberlein13a}
A combination of an fRG flow for $\Lam > \Lam_c$ with a 
mean-field treatment of symmetry-breaking has been formulated
and applied already for a particular fRG version based 
on Wick ordered generating functionals.\cite{reiss07}
However, for calculations beyond the lowest-order truncation,
another fRG version, which is based on the effective action, 
\cite{wetterich93,salmhofer01} turned out to be more efficient, 
as it avoids one-particle reducible contributions, and 
self-energy feedback can be implemented easily.

In this paper we derive a consistent combination of the
{\em one-particle irreducible}\/ fRG with MF theory for symmetry
breaking. The resulting scheme differs from the naive idea 
of plugging the effective interaction at scale $\Lam_c$ (or 
slightly above) into the mean-field equations.
We demonstrate the performance of the combined fRG + MF theory 
by computing antiferromagnetic and superconducting order
parameters in the ground state of the repulsive two-dimensional 
Hubbard model, including the possibility of coexistence of
both orders.

%%% Method %%%%%%%%%%%%%%%%%%%%%%%%%%%%%%%%%%%%%%%%%%%%%%%%%%%%%%%

To see how a mean-field treatment of symmetry breaking can be linked
to the fRG flow, we consider the case of superconductivity as a 
prototype.
Fermionic fRG flow equations for spin-singlet superfluids have been
already derived and studied in detail.
\cite{salmhofer04,gersch08,eberlein10,eberlein13}
In a one-loop truncation with self-energy feedback,\cite{katanin04}
the flow is determined by two coupled flow equations for the 
self-energy $\Sg^{\Lam}$ and the two-particle vertex $\Gam^{\Lam}$, 
respectively. Both quantities contain anomalous components in the 
symmetry-broken regime.

The flow equation for the self-energy (normal and anomalous) is 
given by \cite{fn1}
\begin{equation} \label{floweq_Sg}
 \frac{d}{d\Lam} \Sg_{s_1s_2}^{\Lam}(k) =
 \sum_{s'_1,s'_2} \int_{k'} 
 \Gam^{\Lam}_{s_1s'_1s'_2s_2}(k,k',k',k) \,
 S_{s'_2s'_1}^{\Lam}(k') \, ,
\end{equation}
where 
$S^{\Lam} = 
 \frac{d}{d\Lam} \left. G^{\Lam} \right|_{\Sg^{\Lam} \, {\rm fixed}}$ 
is a scale-derivative of the full propagator $G^{\Lam}$ which acts 
only on its bare part $G^{\Lam}_0$.
The variable $k = (k_0,\bk)$ comprises momentum and Matsubara 
energy, $\int_k$ is a short hand notation for 
$T \sum_{k_0} \int \frac{d^dk}{(2\pi)^d}$, and $s_i = \pm$ 
are indices labeling the two components of Nambu spinors.
$\Sg^{\Lam}_{++}(k) = - \Sg^{\Lam}_{--}(-k) = \Sg^{\Lam}(k)$ is the
normal self-energy, and 
$\Sg^{\Lam}_{+-}(k) = \Sg^{\Lam *}_{-+}(k) = - \Delta^{\Lam}(k)$
the (sign-reversed) gap function.
Note that the vertex enters only with a special choice of momenta
corresponding to zero total momentum (Cooper channel) or zero momentum
transfer (forward scattering).

The flow of the vertex is given by a sum of three distinct one-loop 
contributions.\cite{eberlein13}
It was shown previously that in mean-field models with reduced 
interactions, such as the reduced BCS model, only the channel 
which generates the instability contributes to the vertex flow.
\cite{salmhofer04}
Hence, the other two channels describe fluctuations.
Our strategy is thus to take all contributions to the vertex flow 
into account above the scale for symmetry breaking, but discard the 
fluctuation channels below.
For a singlet superfluid, discarding the fluctuation terms in the 
flow equation for the vertex leads to a simplified flow equation for 
the relevant vertex components $\Gam^{\Lam}_{s_1s_2s_3s_4}(k,k') = 
 \Gam^{\Lam}_{s_1s_2s_3s_4}(k,k',k',k)$,
\begin{eqnarray} \label{floweq_Gam}
 && \frac{d}{d\Lam} \Gam^{\Lam}_{s_1s_2s_3s_4}(k,k') =
 \nonumber \\
 && \sum_{s'_i} \int_p \Gam^{\Lam}_{s_1s'_2s'_3s_4}(k,p) \,
 \dot\Pi^{\Lam}_{s'_1s'_2s'_3s'_4}(p) \,
 \Gam^{\Lam}_{s'_4s_2s_3s'_1}(p,k') \, , \hskip 5mm
\end{eqnarray}
where 
$\Pi^{\Lam}_{s_1s_2s_3s_4}(p) = G^{\Lam}_{s_1s_2}(p) G^{\Lam}_{s_3s_4}(p)$,
and the dot denotes a $\Lam$-derivative.

We denote the scale at which we switch from the full fRG to the mean-field
treatment by $\Lam_{\rm MF}$. Typically $\Lam_{\rm MF}$ will be chosen 
slightly above the critical scale $\Lam_c$.
The full fRG flow for $\Lam > \Lam_{\rm MF}$ yields $\Sg^{\Lam_{\rm MF}}$
and $\Gam^{\Lam_{\rm MF}}$, which pose the initial condition for the
remaining (mean-field) flow for $\Lam < \Lam_{\rm MF}$.
The coupled equations (\ref{floweq_Sg}) and (\ref{floweq_Gam}) for the
self-energy and vertex describing the mean-field flow for
$\Lam < \Lam_{\rm MF}$ can be integrated with arbitrary initial conditions 
at $\Lam = \Lam_{\rm MF}$.
The resulting vertex is determined by a Bethe-Salpeter-type integral 
equation 
\begin{eqnarray} \label{sol_Gam}
 && \Gam^{\Lam}_{s_1s_2s_3s_4}(k,k') =
 \tilde\Gam^{\Lam_{\rm MF}}_{s_1s_2s_3s_4}(k,k') 
 \nonumber \\ 
 && + \sum_{s'_i} \int_p 
 \tilde\Gam^{\Lam_{\rm MF}}_{s_1s'_2s'_3s_4}(k,p) \, 
 \Pi^{\Lam}_{s'_1s'_2s'_3s'_4}(p) \,
 \Gam^{\Lam}_{s'_4s_2s_3s'_1}(p,k') \, , \hskip 5mm
\end{eqnarray}
and the self-energy by a Hartree-type equation of the form
\begin{eqnarray} \label{sol_Sg}
 && \Sg^{\Lam}_{s_1s_2}(k) = \Sg^{\Lam_{\rm MF}}_{s_1s_2}(k)
 \nonumber \\ 
 && + \sum_{s'_1s'_2} \int_{k'} 
 \tilde\Gam^{\Lam_{\rm MF}}_{s_1s'_1s'_2s_2}(k,k')
 \big[ G^{\Lam}_{s'_2s'_1}(k') - G^{\Lam_{\rm MF}}_{s'_2s'_1}(k') 
 \big] \, . \hskip 5mm
\end{eqnarray}
The vertex $\tilde\Gam^{\Lam_{\rm MF}}$ on the right hand sides 
is the {\em irreducible} part of $\Gam^{\Lam_{\rm MF}}$, which can
be determined from the latter via Eq.~(\ref{sol_Gam}) at $\Lam = 
\Lam_{\rm MF}$.
Contributions which are two-particle reducible in the symmetry breaking 
channel are removed in $\tilde\Gam^{\Lam_{\rm MF}}$.
The computation of $\Sg^{\Lam}$ from Eq.~(\ref{sol_Sg}) does not require 
a computation of the vertex for $\Lam < \Lam_{\rm MF}$.

To obtain the physical self-energy and vertex, with all degrees
of freedom integrated, it suffices to solve Eqs.~(\ref{sol_Gam}) 
and (\ref{sol_Sg}) for $\Lam = 0$.
Choosing $\Lam_{\rm MF} \geq \Lam_c$, the vertex $\Gam^{\Lam_{\rm MF}}$
and its irreducible part $\tilde\Gam^{\Lam_{\rm MF}}$ have no
anomalous components, which simplifies the computation considerably.
In particular, the equation for the gap function becomes
\begin{equation} \label{sol_gap}
 \Delta(k) = - \int_{k'} \tilde V^{\Lam_{\rm MF}}(k,k') \,
 F(k') \, ,
\end{equation}
where $F(k)$ is the anomalous propagator, and
$\tilde V^{\Lam_{\rm MF}}(k,k')$ is the irreducible part
of the spin-singlet component of the normal two-particle vertex
\cite{fn2} in the Cooper channel,
\begin{equation} \label{V}
 V^{\Lam_{\rm MF}}(k,k') = 
 \frac{1}{2} \Gam^{\Lam_{\rm MF}}_s(k,-k;-k',k') \, ,
\end{equation}
which is related to the latter by
\begin{eqnarray} \label{V_irr}
 && V^{\Lam_{\rm MF}}(k,k') = \tilde V^{\Lam_{\rm MF}}(k,k') 
 \nonumber \\ 
 && - \int_p \tilde V^{\Lam_{\rm MF}}(k,p) 
 G^{\Lam_{\rm MF}}(p) G^{\Lam_{\rm MF}}(-p) 
 V^{\Lam_{\rm MF}}(p,k') \, . \hskip 7mm
\end{eqnarray}
To compute $\Delta$, one first computes $V^{\Lam_{\rm MF}}$
from the fRG flow, then solves the linear integral equation 
(\ref{V_irr}) for $\tilde V^{\Lam_{\rm MF}}$, and finally the 
gap equation (\ref{sol_gap}).

The fRG + MF procedure described above solves mean-field models
exactly, by construction. For mean-field models, the irreducible 
vertex $\tilde\Gam^{\Lam_{\rm MF}}$ is just the bare vertex for 
any $\Lam_{\rm MF}$.
The integration over momenta and frequencies on the right hand
side of the equation for $\Sg^{\Lam}$ is not restricted by
$\Lam_{\rm MF}$.
This differs from the Wick ordered fRG + MF scheme,\cite{reiss07}
where integrations are restricted by $\Lam_{\rm MF}$ as an 
{\em upper} cutoff.
On the other hand, in that approach the full vertex 
$\Gam^{\Lam_{\rm MF}}$ enters, not only its irreducible part.
However, that scheme, and also its analogue for the one-particle
irreducible fRG~\cite{platt2013}, suffers from systematic errors even for
mean-field models. In particular, the order parameter obtained from solving
the mean-field equations with the full vertex exhibits an artificial
divergence when $\Lam_{\rm MF}$ approaches $\Lam_c$, due to an
overcounting of contributions.

The generalization of the above fRG + MF procedure to other 
instabilities is straightforward. The crucial point is that 
the {\em irreducible} part $\tilde\Gam^{\Lam_{\rm MF}}$ of the 
relevant vertex component has to be inserted as effective
interaction in the mean-field equation for the order parameter.
The computation of $\tilde\Gam^{\Lam_{\rm MF}}$ from
$\Gam^{\Lam_{\rm MF}}$ is done for each instability channel
separately, even in cases of coexistence of order in different
channels.

%%% AF + SC %%%%%%%%%%%%%%%%%%%%%%%%%%%%%%%%%%%%%%%%%%%%%%%%%%

To illustrate the performance of the fRG + MF theory in a 
situation of competing instabilities, we now present an
application to the two-dimensional Hubbard model.
The model is well-known for its intriguing competition
between antiferromagnetism and superconductivity.
\cite{scalapino12}
Indeed the fRG flow of the vertex generically diverges either 
in the antiferromagnetic or in the $d$-wave pairing channel in
that model.\cite{metzner12}
Hence we allow for antiferromagnetic and superconducting order,
including the possibility of coexistence.
Although the antiferromagnetic wave vector may deviate from 
$(\pi,\pi)$, we consider only the case of conventional Ne\'el 
order for simplicity. 

The effective interaction for singlet pairing is given by
Eq.~(\ref{V}), and its irreducible part by Eq.~(\ref{V_irr}).
Similarly, the effective interaction triggering antiferromagnetism
is given by
\begin{equation}
 U^{\Lam_{\rm MF}}(k,k') = \sum_{\sg'=\up,\down}
 s_{\sg'} \Gam^{\Lam_{\rm MF}}_{\up\sg'\sg'\up}(k+Q,k',k'+Q,k)
 \, ,
\end{equation}
where $s_{\up},s_{\down} = \pm 1$, and $Q = (0,\bQ)$ with the
antiferromagnetic wave vector $\bQ = (\pi,\pi)$.
Its irreducible part is obtained from the integral equation
\begin{eqnarray} \label{U_irr}
 && U^{\Lam_{\rm MF}}(k,k') = \tilde U^{\Lam_{\rm MF}}(k,k')
 \nonumber \\ 
 && + \int_p \tilde U^{\Lam_{\rm MF}}(k,p) 
 G^{\Lam_{\rm MF}}(p) G^{\Lam_{\rm MF}}(p+Q) 
 U^{\Lam_{\rm MF}}(p,k') \, . \hskip 5mm
\end{eqnarray}

So far, the formalism allows for dynamical (frequency dependent) 
effective interactions and order parameters.
In this first application, we will discuss only the static
mean-field theory obtained from the static (zero frequency)
effective interactions $U^{\Lam_{\rm MF}}_{\bk\bk'}$ and 
$V^{\Lam_{\rm MF}}_{\bk\bk'}$.
We will also discard normal self-energy contributions.
The superconducting and antiferromagnetic order parameters
are then defined as gap functions in the usual form \cite{reiss07}
\begin{eqnarray}
 \Delta^{\rm SC}_{\bk} &=& \int_{\bk'} 
 \tilde V^{\Lam_{\rm MF}}_{\bk\bk'} \bra p_{\bk'} \ket \, , \\
 \Delta^{\rm AF}_{\bk} &=& \frac{1}{2} \int_{\bk'} 
 \tilde U^{\Lam_{\rm MF}}_{\bk\bk'} \bra m_{\bk'} \ket \, ,
\end{eqnarray}
where $p_{\bk} = a_{\bk\up} a_{-\bk\down}$ is the Cooper
pair annihilation operator, 
$m_{\bk} = a^{\dag}_{\bk\up} a_{\bk+\bQ,\up} -
 a^{\dag}_{\bk\down} a_{\bk+\bQ,\down}$ is the operator for
staggered magnetization, and
$\int_{\bk}$ is an abbreviation for $\int \! \frac{d^2k}{(2\pi)^2}$.
We choose the phase of the superconducting order parameter such
that $\Delta^{\rm SC}_{\bk}$ is real.
A mean-field decoupling of the reduced effective interactions
yields the mean-field Hamiltonian
\begin{eqnarray}
 H_{\rm MF} &=& H_0 +
 \int_{\bk} \Delta^{\rm AF}_{\bk} 
 \left( m_{\bk} - {\textstyle\frac{1}{2}} \bra m_{\bk} \ket \right) 
 \\
 &+&  \int_{\bk} \Delta^{\rm SC}_{\bk} 
 \left( p_{\bk} + p^{\dag}_{\bk} - 
 {\textstyle\frac{1}{2}} \bra p_{\bk} + p^{\dag}_{\bk} \ket \right) 
 \, ,
\end{eqnarray}
where $H_0 = \int_{\bk} \eps_{\bk} n_{\bk}$ is the kinetic
energy. For the Hubbard model with nearest and next-to-nearest
neighbor hopping on a square lattice, the dispersion relation is
$\eps_{\bk} = -2t (\cos k_x + \cos k_y) - 4t' \cos k_x \cos k_y$.
% \tilde V = U and \tilde U = - U for Hubbard in MFT.

The mean-field Hamiltonian can be diagonalized by a Bogoliubov 
transformation,\cite{reiss07} and the resulting gap equations 
can be solved numerically by iteration.
Occasionally two distinct locally stable solutions of the gap
equations are found. One then has to compute the corresponding
free energies to discriminate globally stable from metastable
states.
In case of coexistence of antiferromagnetism and superconductivity,
an additional triplet pairing with pair momentum $(\pi,\pi)$ is
generically generated.\cite{psaltakis83,murakami98,kyung00}
However, its feedback on the main order parameters is very weak,
\cite{reiss07} so that we can safely discard this additional
order parameter in the computation of $\Delta^{\rm AF}_{\bk}$
and $\Delta^{\rm SC}_{\bk}$.

%%% Results %%%%%%%%%%%%%%%%%%%%%%%%%%%%%%%%%%%%%%%%%%%%%%%%%%%%%%%%

We now show and discuss results for the magnetic and superconducting
order parameters in the ground state of the hole-doped Hubbard model 
with a small next-to-nearest neighbor hopping $t'/t=-0.15$ and a 
moderate Hubbard interaction $U/t = 3$.
The fRG flow has been computed with a static vertex parametrized
via a decomposition in charge, magnetic and pairing channels,
\cite{karrasch08,husemann09} with $s$-wave and $d$-wave form 
factors as described in Ref.\ \onlinecite{eberlein13a}.
The scale $\Lam_{\rm MF}$ was fixed by the condition that the modulus
of one of the coupling functions parametrizing the vertex reaches the
maximal value $50t$.
With this criterion $\Lam_{\rm MF}$ is typically less than 10 percent 
above $\Lam_c$.   
For the computation of $\tilde\Gam^{\Lam_{\rm MF}}$ and the solution 
of the gap equations, the momentum dependence was discretized by 
partitioning the Brillouin zone in 100 patches.

\begin{figure}% [htb]
\begin{center}
\includegraphics[width=8cm]{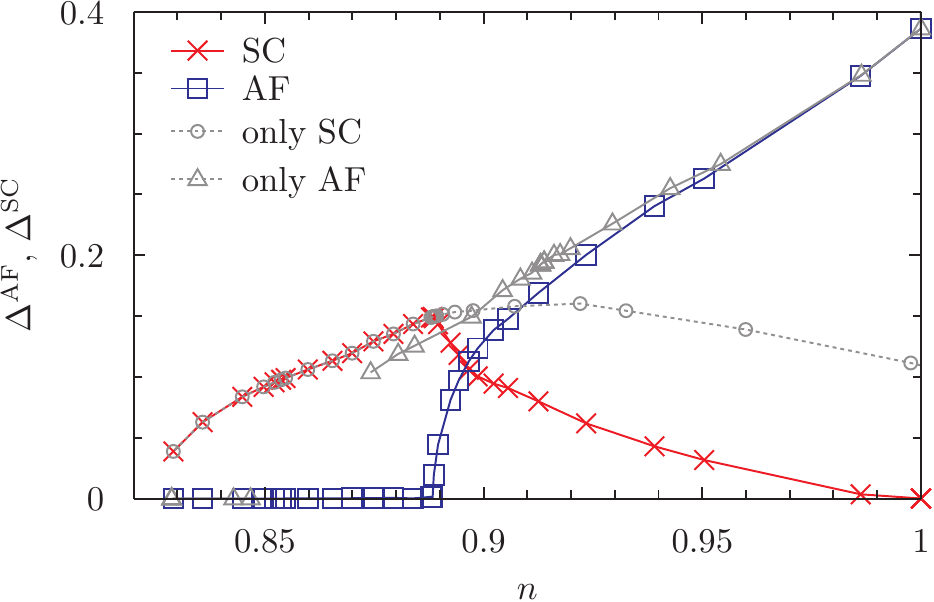}
\caption{(Color online) Amplitudes of antiferromagnetic and
 superconducting gap functions in the ground state of the
 two-dimensional Hubbard model as a function of density, for
 $U/t=3$ and $t'/t=-0.15$. Results from a coupled solution of
 the magnetic and superconducting gap equations with partial 
 coexistence of orders are compared to purely magnetic and
 purely superconducting solutions. The amplitudes are plotted
 in units of $t$.}
\end{center}
\end{figure}
In Fig.~1 we show results for the amplitudes of the antiferromagnetic
and superconducting gap functions, 
$\Delta^{\rm AF} = \max_{\bk} \Delta^{\rm AF}_{\bk}$ and 
$\Delta^{\rm SC} = \max_{\bk} \Delta^{\rm SC}_{\bk}$, 
as a function of the electron density.
The coupled solution of both gap equations exhibits an extended 
region where magnetic and superconducting order coexist. In the
major part of that region the pairing gap is smaller than the
magnetic gap. Here superconductivity is a secondary instability 
within the antiferromagnetic phase, which naturally occurs as a
Cooper instability of electrons near the reconstructed Fermi 
surface confining hole pockets in the antiferromagnetic state.
The pairing gap decreases rapidly as the pockets shrink upon 
approaching half-filling.
Magnetic order vanishes at a critical density $n_c^{\rm AF}$
situated slightly above Van Hove filling. Below that density the
state is purely superconducting. The magnetic transition is 
{\em continuous}\/ such that $n_c^{\rm AF}$ is a quantum critical 
point.\cite{qcp}
Fig.~1 also shows results for the gap amplitudes as obtained from 
solutions of the individual gap equations with either magnetic
or superconducting order. A comparison with the coupled solution
confirms that the two order parameters compete with each other.
In particular, superconductivity is strongly suppressed by 
antiferromagnetism. In the absence of superconductivity, the
antiferromagnetic regime extends to lower densities and terminates
at a first order transition accompanied by a density jump,
which opens a density window where no homogeneous solution
exists.

For densities below $n=0.95$, the two-particle vertex diverges 
actually at incommensurate wave vectors, indicating a leading
instability toward incommensurate antiferromagnetic order.
\cite{eberlein13a} The resulting ground state is probably an
incommensurate spin density wave state coexisting with
superconductivity. Such states can also be treated by the
fRG + MF theory. Since mean-field equations for incommensurate
magnetic order are more involved, we leave this extension for 
future studies.
For parameters where pairing is the leading instability, the
results for $\Delta^{\rm SC}$ are very close to those from a
full fRG calculation,\cite{eberlein13a} which indicates that
the fluctuations below the scale $\Lam_c$ have indeed limited
impact on the size of the ground state order parameter.

\begin{figure} % [htb]
\begin{center}
\includegraphics[width=8cm]{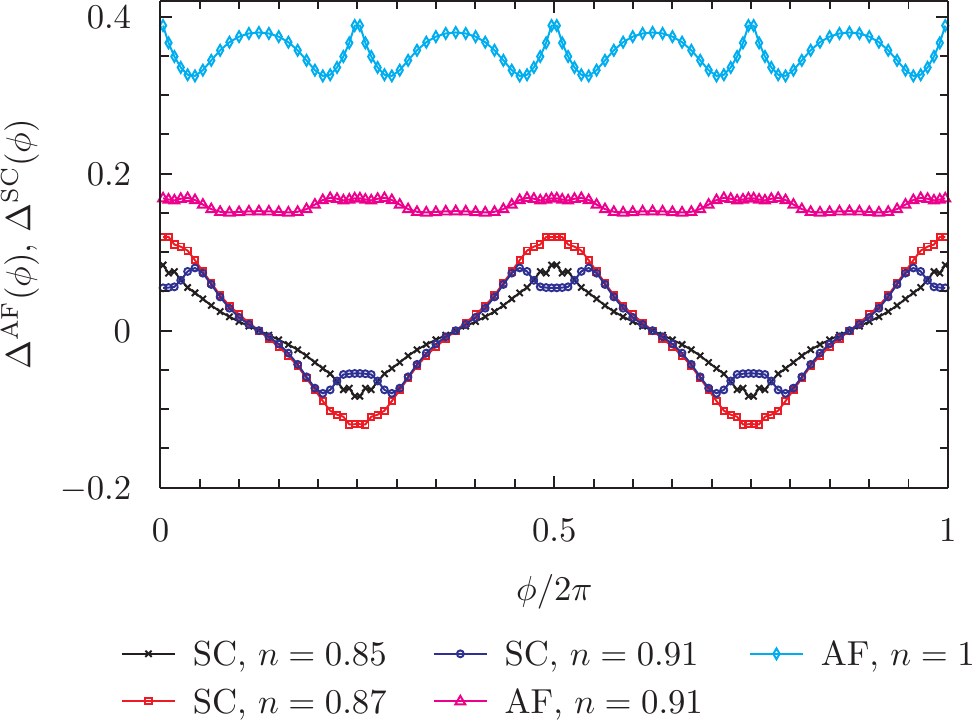}
\caption{(Color online) Momentum dependence of the 
 antiferromagnetic and superconducting gap functions 
 for various choices of the density.
 The momentum dependence is parametrized by the angle
 $\phi$ between $\bk$ and the $k_x$-axis, where 
 $\Delta^{\rm AF}_{\bk}$ is evaluated with $\bk$ on the
 Umklapp surface ($|k_x \pm k_y| = \pi$) and 
 $\Delta^{\rm SC}_{\bk}$ with $\bk$ on the Fermi surface.
 The model parameters are the same as in Fig.~1.}
\end{center}
\end{figure}
The momentum dependence of the gap functions is shown in Fig.~2.
The antiferromagnetic gap $\Delta^{\rm AF}_{\bk}$ exhibits only
a moderate modulation around a constant.
The superconducting gap $\Delta^{\rm SC}_{\bk}$ obeys the 
expected $d_{x^2-y^2}$ symmetry, but with visible deviations
from the simple $\cos k_x - \cos k_y$ form.
In the coexistence regime the (global) extrema of 
$\Delta^{\rm SC}_{\bk}$ are shifted away from the axial 
directions, as a natural consequence of the Fermi surface
truncation in the antinodal region.

The above results for the gap functions agree qualitatively with 
those obtained previously from the Wick ordered fRG + MF theory.
\cite{reiss07} However, the suppression of $\Delta^{\rm SC}_{\bk}$
by antiferromagnetic order was stronger in that work.
A relatively broad coexistence of antiferromagnetism and 
superconductivity as found here has also been obtained at stronger
interactions by embedded quantum cluster methods.
\cite{lichtenstein00,capone06,aichhorn06}
The results for the order parameters depend to some extent on the 
precise choice of $\Lam_{\rm MF}$, but much less than in the Wick 
ordered fRG + MF approach.

%%% Summary %%%%%%%%%%%%%%%%%%%%%%%%%%%%%%%%%%%%%%%%%%%%%%%%%%

In summary, we have derived an efficient and unbiased method
for computing order parameters in correlated electron systems
with competing instabilities. 
Charge, magnetic and pairing fluctuations above the energy scale 
of symmetry breaking are taken into account by an fRG flow, 
while the formation of order below that scale is treated in 
mean-field theory.
The effective interaction entering the mean-field equations is
given by the irreducible part of the two-particle vertex.
The method captures fluctuation driven instabilities such as
$d$-wave superconductivity in two-dimensional electron systems.
It can deal with any order parameter based on a bilinear fermionic 
expectation value.
As a first application we have studied the competition between
antiferromagnetism and superconductivity in the two-dimensional
Hubbard model.
An interesting extension would be the computation of incommensurate 
magnetic order, in possible coexistence with superconductivity,
which is very hard to study by other methods.
More generally, competing instabilities in complex multi-band 
systems offer a wide field of fruitful applications for the 
fRG + MF theory.

%\vskip 5mm

%%%%%%%%%%%%%%%%%%%%%%%%%%%%%%%%%%%%%%%%%%%%%%%%%%%%%%%%%%%%%%%%

\begin{acknowledgments}
We would like to thank F.~Benitez, O.~Gunnarsson, C.~Honerkamp,
M.~Salmhofer, and H.~Yamase for valuable discussions.
J.W.\ was supported by the joint MPG/CAS doctoral promotion
program.
Support from the DFG research group FOR 723 is also gratefully
acknowledged.
\end{acknowledgments}

%%%%%%%%%%%%%%%%%%%%%%%%%%%%%%%%%%%%%%%%%%%%%%%%%%%%%%%%%%%%%%%%%%%%


\begin{thebibliography}{99}
\bibitem{fn1} We use the same notation and sign conventions as in
 Ref.\ \onlinecite{eberlein13}.
\bibitem{fn2} For spin-rotation invariant systems, the normal 
 two-particle vertex can be decomposed in spin singlet and
 triplet components as
 $\Gam_{\sg_1\sg_2\sg_3\sg_4}(k_1,k_2;k_3,k_4) =
  \Gam_s(k_1,k_2;k_3,k_4) S_{\sg_1\sg_2\sg_3\sg_4} +
  \Gam_t(k_1,k_2;k_3,k_4) T_{\sg_1\sg_2\sg_3\sg_4}$, where
 $S_{\sg_1\sg_2\sg_3\sg_4} = 
  \frac{1}{2} (\delta_{\sg_1\sg_4} \delta_{\sg_2\sg_3} -
  \delta_{\sg_1\sg_3} \delta_{\sg_2\sg_4} )$ and
 $T_{\sg_1\sg_2\sg_3\sg_4} = 
  \frac{1}{2} (\delta_{\sg_1\sg_4} \delta_{\sg_2\sg_3} +
  \delta_{\sg_1\sg_3} \delta_{\sg_2\sg_4} )$.
\end{thebibliography}
\end{document}